\documentclass{aa} 
\usepackage{graphicx} 

\usepackage[fleqn]{amsmath}
\usepackage{txfonts}
\usepackage{natbib}
\bibpunct{(}{)}{;}{a}{}{,}

\def\fduga{\ensuremath{f_\mathrm{d/g}}}
\def\um{\ensuremath{\mu\mathrm{m}}}

\begin{document}

\title{Deuterated H$_3^+$ in proto-planetary disks}

\author{Cecilia Ceccarelli\inst{1}\& Carsten Dominik\inst{2}}

\offprints{Cecilia.Ceccarelli@obs.ujf-grenoble.fr}

\institute{Laboratoire d'Astrophysique, Observatoire de Grenoble - 
BP 53, F-38041 Grenoble cedex 09, France 
\and 
Sterrenkundig Instituut ``Anton Pannekoek'', Kruislaan 403, NL-1098SJ
Amsterdam, The Netherlands}
\date{Received 4 Mars 2005; accepted 6 June 2005}

\titlerunning{Deuterated H$_3^+$ in proto-planetary disks}

\authorrunning{Ceccarelli \& Dominik}

\abstract{Probing the gas and dust in proto-planetary disks is central
  for understanding the process of planet formation.  In disks
  surrounding solar type protostars, the bulk of the disk mass resides
  in the outer midplane, which is cold ($\leq$20 K), dense ($\geq
  10^7$ cm$^{-3}$) and depleted of CO.  Observing the disk midplane
  has proved, therefore, to be a formidable challenge. Ceccarelli et
  al. (2004) detected H$_2$D$^+$ emission in a proto-planetary disk
  and claimed that it probes the midplane gas.  Indeed, since all
  heavy-elements bearing molecules condense out onto the grain
  mantles, the most abundant ions in the disk midplane are predicted
  to be H$_3^+$ and its isotopomers. In this article, we carry out a
  theoretical study of the chemical structure of the outer midplane of
  proto-planetary disks. Using a self-consistent physical model for
  the flaring disk structure, we compute the abundances of H$_3^+$ and
  its deuterated forms across the disk midplane.  We also provide the
  average column densities across the disk of H$_3^+$, H$_2$D$^+$,
  HD$_2^+$ and D$_3^+$, and line intensities of the ground transitions
  of the ortho and para forms of H$_2$D$^+$ and HD$_2^+$
  respectively. We discuss how the results depend on the cosmic ray
  ionization rate, dust-to-gas ratio and average grain radius, and
  general stellar/disk parameters.  An important factor is the poorly
  understood freeze-out of N$_2$ molecules onto grains, which we
  investigate in depth.  We finally summarize the diagnostic values of
  observations of the H$_3^+$ isotopomers.  \keywords{ISM: abundances
  -- ISM: molecules -- Stars: formation --ISM: } }

\maketitle

%
%-----------------------------------------------------------------------

\section{Introduction}

Proto-Planetary disks are the sites of planet formation.  Their
physical, dynamical and chemical structure and evolution determine if,
when, how, where and what planets form.  Two very
important parameters that are difficult to determine are the gas mass
of the disk, and the ionization degree.  Critical questions that need
to be answered by observations are: what is the evolution of the
gaseous component of the disk, in particular with respect to the dusty
component? Is it dispersed before or after dust coagulates into
planetesimals and/or rocky planets?  At what radius?  By what process?
At the same time, theory predicts that the accretion in the disk is
regulated by its ionization degree (Balbus \& Hawley 1998, Gammie
1996). So the questions here are: what is actually the measured
ionization degree across the disk?  What ionizes the gas? Cosmic rays
and/or X-rays?  Where do the two effects balance each other, if they
do?

Observationally answering to those questions is all but an easy task,
especially in solar type systems.  This is because the bulk of the
disk mass resides in the outer midplane, which is cold and dense. As a
consequence, all heavy-bearing molecules freeze-out onto the grain
mantles, and disappear from the gas phase where they could be
observed.  So the first difficulty in the study of the outer disk
midplane is to find probes of it.  Last year, Ceccarelli et al. (2004)
detected abundant H$_2$D$^+$ in the proto-planetary disk which
surrounds the solar type protostar DM Tau.  They claimed that
H$_2$D$^+$ probes the cold outer midplane and, in addition, its
abundance is a direct measure of the ionization degree.  The reason
behind this claim is the peculiar chemistry of the H$_2$D$^+$ ion.
The basic idea is that in cold and dense gas, where CO and all
heavy-bearing molecules freeze-out into the grain mantles, two things
happen: first, only H$_2$ and the ions from this molecule, namely
H$_3^+$ and its isotopomers, remain in the gas phase; second, the
molecular deuteration is dramatically enhanced, up to having
H$_2$D$^+$/H$_3^+$ larger than unity (Caselli et al. 2003).
Therefore, Ceccarelli et al. concluded that a) H$_2$D$^+$ line
emission probes the gas disk midplane, and b) the H$_2$D$^+$ abundance
measures the ionization degree there.

In the present article, we examine the chemistry of the deuterated
forms of H$_3^+$ in proto-planetary disks, with the goal of exploring,
on a solid theoretical basis, the exact diagnostic value of the
H$_2$D$^+$ observations. Besides, the present study concerns also the
other H$_3^+$ isotopomers, HD$_2^+$ and D$_3^+$.  To accomplish the
goal, we develop a chemical model of the outer disk midplane, focused
in particular on the H$_3^+$ deuteration chemistry.  The physical
model that describes the disk computes self-consistently the
temperature and density profiles for a given disk mass and star
luminosity (Dullemond et al. 2001; Dullemond \& Dominik 2004). The
chemical model is based on what has been understood from the studies
of molecular deuteration in pre-stellar-cores and protostars, both
observationally (see e.g.  the review by Ceccarelli 2004) and
theoretically (e.g. Roberts, Millar \& Herbst 2003, 2004; Walmsley,
Flower \& Pineaut des Forets 2004, Flower, Pineaut des Forets \&
Walmsley 2004).  Particular emphasis is devoted to the role of dust
grains, and the effect of dust coagulation/fragmentation on the disk
midplane chemical structure.

Aiming to give observable predictions, besides to provide the
abundances of the H$_3^+$ isotopomers across the disk, and the average
column density of each species, we compute the intensities of the four
ground state lines from the ortho and para forms of H$_2$D$^+$ and
HD$_2^+$ respectively.  We carry out a wide parameter study, and
explore the dependence of our results on three major parameters: the
dust-to-gas ratio, the cosmic rays ionization rate, and the dust grain
average sizes.  Besides, we also discuss how the results depend on the
basic properties of the star-disk system, namely the star luminosity,
disk age, mass and radius.  Finally, we discuss the case in which
N$_2$ disappears from the gas phase simultaneously with CO, which is
not what has been observed so far, but what it would be expected based
on the laboratory measurements of the N$_2$ and CO binding energies
(e.g. Oberg et al. 2005).

The article is organized as follows. In \S \ref{sec:model} we develop
our model. In \S \ref{sec:results} we report the model predictions of
a standard case, and as function of the parameters of the model.  In \S
\ref{sec:discussion} we discuss the diagnostic values of the
observable quantities (line intensities and column densities).
Finally, \S \ref{sec:conclusions} summarizes the content of the
article.

\section{The model}\label{sec:model}

In this section we describe the model that we developed to calculate
the abundance profiles and the average column densities of H$_3^+$ and
its isotopomers across the disk, and the line emission of the two
deuterated forms of H$_3^+$ which have observable ground state
rotational transitions, H$_2$D$^+$ and HD$_2^+$.

\subsection{Disk structure \label{phys}}

\begin{figure}[tb]
\includegraphics[angle=0,width=1.\columnwidth]%
%{../chemistry/output/struct_standard.ps}
{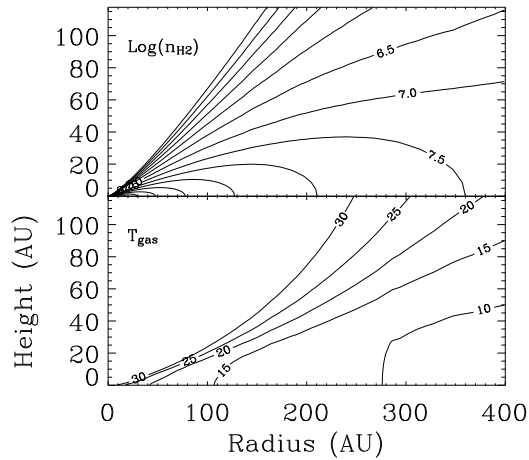}
\caption{The density of H$_2$ molecules (top panel), and the local
  dust temperature in the standard disk model (bottom panel),
  described in \S \ref{sec:standard-model}.}
\label{fig:struct}
\end{figure}

For the model calculations we use a model of passively irradiated
hydrostatic flaring circumstellar disks (Dullemond et al 2001;
Dullemond \& Dominik 2004).  This model computes the structure (i.e.
the density and temperature distribution) in a selfconsistent way.  In
the center of the disk, a low-mass star is located with a mass of
$M_\star$ and a luminosity of $L_{\star}$.  Around the star we
distribute $M_\mathrm{disk}$ of material in a disk ranging from
0.1\,AU to $R_\mathrm{disk}$, with a surface density powerlaw
$\Sigma\propto r^{-1}$, implying that the disk mass per unit radius is
constant, i.e. both inner and outer disk contain significant amounts
of mass.  The disk contains dust at a mass fraction of $\fduga$ which
we assume to be fully mixed with the gas.  While in reality, there
probably exists a distribution of grain sizes in the disk, we choose a
single grains size for the present calculation.  This allows us in a
simple way to study the effects of grain size.  The structure of the
disk is then computed by iterating between a 1+1D continuum radiative
transfer which computes the dust temperature in the entire disk, and a
hydrostatic equilibrium code which computes the vertical density and
pressure distribution under the assumption that the gas and dust
temperatures are equal.  For the technical details of the modeling
procedure we refer to Dullemond et al (2001).  A typical temperature
and density profile across the disk (our standard case described in \S
\ref{sec:standard-model}) is shown in Fig.\ref{fig:struct}.  The disk
is flaring as can be seen from the upwards-curved temperature
contours.  We anticipate that heavy-elements bearing molecules will
freeze-out, and therefore H$_3^+$ deuterium fractionation will be
significant at about 25 K (approximatively the CO condensation
temperature for the involved densities) and below.  These conditions
are only fulfilled in the outer disk, approximately outside of 20
AU. The low temperature region becomes geometrically thick at large
distances from the star.  For example, at a distance of 300 AU, the 15
K contour reaches a height of 60 AU.  Typical densities in this region
are $n_\mathrm{H_2}=10^7$ cm$^{-3}$ and above, up to
$n_\mathrm{H_2}=10^{8.5}$ cm$^{-3}$ in the innermost parts of the
outer disk, near 100\,AU.

\subsection{Chemistry of deuterated H$_3^+$}\label{sec:chem-deut-h_3+}

Complex chemical networks are often necessary to compute the
abundances of molecules in interstellar environments.  In particular
in regions with a rich chemistry, it is difficult to predict which are
the important reactions leading to and from a certain molecule
(e.g. Semenov, Wiebe \& Henning 2004).  However, in the cold and
metal-depleted regions in protostellar cores and circumstellar disks,
the chemical network is limited and a much simpler treatment is
possible.  Also, if the chemical timescales involved are short, a
steady state solution of the chemistry is often appropriate.  In the
following we address the most important processes leading to the
formation of deuterated H$_3^+$ and develop a set of equations for the
steady state solution of this network.

In gas under standard molecular cloud conditions, H$_3^+$ ions are
formed with a rate $\zeta$ by the ionization of H$_{2}$ (and He) due to cosmic
rays, and destroyed by the reactions with all neutral molecules and
atoms in the gas, and by recombination reactions with electrons and
grains.  The reaction H$_3^+$ with H$_2$ -- the most abundant species
in molecular gas -- returns H$_3^+$ and therefore has no net effect on
the H$_3^+$ abundance.  However, the reaction with HD, the most
abundant H$_2$ isotopomer, forms H$_2$D$^+$:
\begin{equation}\label{h3+}
\mathrm{H_3^+ + HD \rightarrow H_2D^+ + H_2}
\end{equation}
with a rate coefficient $k_1$ (see Table 1).
The backward reaction is endothermic with an energy barrier of about
220\,K.  At low temperatures, it is inefficient, making the 
reaction \ref{h3+} the effective route to H$_2$D$^+$ formation.  When
considering the abundances in cold molecular gas, the most important
molecules which cause destruction of H$_3^+$ are: HD, CO
(the second most abundant molecule after H$_{2}$), and N$_2$ (see \S
\ref{sec:co-n_2-freezing}).  Also recombination with grains (see
\S~\ref{sec:recombination-grains}) is an important process.  Equating
H$_3^+$ formation and destruction rates, and ignoring the destruction
by molecules/atoms less abundant than CO and N$_2$, the equilibrium abundance
ratio between the H$_3^+$ and H$_2$D$^+$ is given by the following
equation:
\begin{equation}\label{eq:ratio1}
\mathrm{%
\frac{H_2D^+}{H_3^+}}=
\frac{2 \cdot [\mathrm{D}] k_{1}}{%
k_\mathrm{rec1}x_\mathrm{e} + 
k_\mathrm{CO}x_\mathrm{CO} +
k_\mathrm{N_2}x_\mathrm{N_2} +
k_\mathrm{gr}x_\mathrm{gr} +
2\cdot k_{-1}[\mathrm{D}] +
2k_{2}[\mathrm{D}]
}
\end{equation}
where the rate coefficients are defined in Table \ref{tab:reactions},
$x_e$ is the electronic abundance, and [D] is the elemental abundance
of deuterium relative to H nuclei, equal to $1.5\times10^{-5}$ (Lynsky
2003).

\begin{table}
\def\r#1{$\mathrm{#1}$}
\def\ra{$\rightarrow$}
\def\lra{$\leftrightarrow$}
\small
\begin{tabular}[tb]{lllllrr}
Reaction & & & rate & $\gamma$ & $\alpha$  & $\beta$ \\
         & & &      & cm$^3$s$^{-1}$ &     & K\\\hline
\r{H_3^+ + HD}   & \lra & \r{H_2D^{+}+H_2}& $k_1,k_{-1}$      & 1.7(--9 ) &   0.0 & 220  \\
\r{H_3^+ + CO}   & \ra  &                & $k_\mathrm{CO}$    & 6.6(--10) & --0.5 &      \\
\r{H_3^+ + e^-}  & \ra  &                & $k_\mathrm{rec0}$  & 6.8(--8)  & --0.5 &      \\
\r{H_3^+ + g}    & \ra  &                & $k_\mathrm{gr1}$   & & &                    \\
\r{H_2D^+ + HD}  & \lra & \r{HD_2^{+}+H_2}& $k_2,k_{-2}$      & 8.1(--10) &   0.0 & 187  \\
\r{H_2D^+ + CO}  & \ra  &                & $k_\mathrm{CO}$    & 6.6(--10) & --0.5 &      \\
\r{H_2D^+ + e^-} & \ra  &                & $k_\mathrm{rec1}$  & 6.0(--8)  & --0.5 &      \\
\r{H_2D^+ + g}   & \ra  &                & $k_\mathrm{gr2}$   & & &                    \\
\r{HD_2^+ + HD}  & \lra & \r{D_3^{+}+H_2}& $k_3,k_{-3}$       & 6.4(--10) &   0.0 & 234  \\
\r{HD_2^+ + CO}  & \ra  &                & $k_\mathrm{CO}$    & 6.6(--10) & --0.5 &      \\
\r{HD_2^+ + e^-} & \ra  &                & $k_\mathrm{rec2}$  & 6.0(--8)  & --0.5 &      \\
\r{HD_2^+ + g}   & \ra  &                & $k_\mathrm{gr3}$   & & & \\
\r{D_3^{+} + CO} & \ra  &                & $k_\mathrm{CO}$    & 6.6(--10) & --0.5 &      \\
\r{D_3^+ + e^-}  & \ra  &                & $k_\mathrm{rec3}$  & 2.7(--8)  & --0.5 &      \\
\r{D_3^+ + g}    & \ra  &                & $k_\mathrm{gr4}$   & & & \\\hline
\end{tabular}\newline
Notes: $a(-b)$ implies $a\cdot 10^{-b}$
\caption{\label{tab:reactions}Reactions and reaction constants
  important for the abundance of H$_3^+$ and its isotopomers.  The
  parameters $\gamma$, $\alpha$, and $\beta$ determine the rate
  coefficient at temperature $T$ through
  $k=\gamma\cdot(T/300K)^{\alpha}\exp\{-\beta/T\}$ and are taken from
  Roberts et al. (2004) and the UMIST database. A recent study
  suggests a lower value of the reaction rate of H$_3^+$ with HD
  (Gerlich, Herbst \& Roueff 2002). However, the study needs
  confirmation, so we preferred to stick on the old value, which also
  offers an easier comparison with previous similar models. In the
  Table we only detail the products of those reactions producing
  H$_3^+$ and its isotopomers.  The destruction reactions involving
  CO, e$^-$ and grains may have a variety of products.  In order to
  compute the rate coefficients for the reactions with CO and e$^-$,
  we added the $\gamma$ parameters of all relevant reactions from the
  UMIST database.  The reaction rates with grains are described in
  \S~\ref{sec:recombination-grains}.}
\end{table}
As noticed by other authors (e.g.  Caselli et al. 2003), the
H$_2$D$^+$/H$_3^+$ ratio, given by Equation \ref{eq:ratio1}, can be
larger than unity if the gas is cold and very depleted of CO and
N$_2$.  In the limit of very cold and completely depleted gas, it
reaches ${k_{1}}/{k_{2}}\sim2$ (Walmsley et al. 2004).

Just like H$_3^+$, H$_2$D$^+$ is destroyed by reacting with molecules,
atoms, electrons, and grains.  In analogy with H$_3^+$, the
interaction of H$_2$D$^+$ with HD leads to the formation of HD$_2^+$,
whereas the reaction with H$_2$ is endothermic and therefore
suppressed at low temperatures.
\begin{equation}\label{h2d+}
\mathrm{H_2D^+ + HD \rightarrow HD_2^+ + H_2} \quad.
\end{equation}
As in the case of reaction (\ref{h3+}), the reverse reaction of
(\ref{h2d+}) is endothermic, and therefore inhibited at low
temperatures.  H$_2$D$^+$ is therefore destroyed by reactions with HD,
CO, N$_2$, electrons, and grains, and forms HD$_2^+$.  In the same way, the
triply deuterated form of H$_3^+$ is formed by the successive reaction
of HD$_2^+$ with HD:
\begin{equation}\label{d3+}
\mathrm{HD_2^+ + HD \rightarrow D_3^+ +H_2}
\end{equation}
and destroyed by the reaction with CO, N$_2$, grains and electrons.  Now the
reaction D$_3^+$ + HD just exchange the deuterium atoms, and D$_3^+$
is therefore the dead-end of the chain.  As in the case of Equation
\ref{eq:ratio1}, the abundance ratios of HD$_2^+$ and D$_3^+$ with
respect to H$_2$D$^+$ and HD$_2^+$ respectively can be derived by
equating formation and destruction rates:
\begin{align}
\mathrm{%
\frac{HD_2^+}{H_2D^+}}&=
\frac{2 \cdot [\mathrm{D}] k_{2}}%
{k_\mathrm{rec2}x_\mathrm{e} + 
k_\mathrm{CO}x_\mathrm{CO} + 
k_\mathrm{N_2}x_\mathrm{N_2} + 
k_\mathrm{gr}x_\mathrm{gr} +
2\cdot k_{-2}[\mathrm{D}] + 
2k_{3}[\mathrm{D}]}
\label{eq:ratio2}\\
\mathrm{%
\frac{D_3^+}{HD_2^+}}&=
\frac{2 \cdot [\mathrm{D}] k_{3}}%
{k_\mathrm{rec3}x_\mathrm{e} +
k_\mathrm{CO}x_\mathrm{CO} +
k_\mathrm{N_2}x_\mathrm{N_2} + 
k_\mathrm{gr}x_\mathrm{gr} +
2k_{-3}[\mathrm{D}]}
\label{eq:ratio3}
\end{align}

One can easily see that the HD$_2^+$/H$_2$D$^+$ ratio can reach unity.
In the limiting case of very low temperatures, extreme CO and N$_2$
depletion and low ionization, it reaches a value of
${k_{2}}/{k_{3}}\sim1.3$.  On the other hand, the abundance of D$_3^+$
is only limited by the electron abundance.  In extreme conditions --
cold and heavily CO and N$_2$ depleted gas -- D$_3^+$ can be the
dominant charge carrier.

Table~\ref{tab:reactions} summarizes the reactions involving H$_3^+$
and its isotopomers, along with the rates used in this study.  The
reaction rates have been taken from Roberts et al. (2003, 2004),
except for recombination reactions on grains which we discuss in \S
\ref{sec:recombination-grains}.

In practice, the adopted chemical network is a small subset of the
extensive chemical networks implemented by Roberts et al. (2003, 2004)
or Walmsley et al.  (2004) or Flower et al. (2004).  Our goal here is
to study the behavior of the H$_3^+$ and its isotopomers, which can
indeed be described by the few equations discussed above.  This sets
the limit of applicability of the present model to regions where CO
and N$_2$ are the only molecules affecting the H$_3^+$ chemistry.
Furthermore, we did not treat the ortho and para forms independently,
but used, when necessary, the results by Walmsley et al. (2004)
and Flower et al. (2004).

With this reduced chemical network, we are able to reproduce the
results by Roberts et al. (2003, 2004), Walmsley et al. (2004) and
Flower et al. (2004) accurately as far as the abundances of the
deuterated forms of H$_3^+$ are concerned.

\subsection{CO and N$_2$ freeze-out}\label{sec:co-n_2-freezing}
The CO and N$_2$ abundances are obviously critical parameters in the
deuteration of H$_3^+$.  CO is known to freeze-out onto the grains
mantles at large enough densities ($\geq 10^5$ cm$^{-3}$) and low
 temperatures ($\leq 25$ K) (e.g. Caselli et al.  1999; Bergin et al.
2001; Bacmann et al. 2002; Tafalla et al. 2002).  Recent laboratory
experiments show that the CO binding energy depends on the matrix
where CO is embedded (e.g.  Collings et al. 2003; Fraser et al. 2004;
Oberg et al. 2005).  It varies from $\sim$ 840 K in CO-CO ices
(average of the values measured by Collings et al. 2003 and Oberg et
al. 2005), to 885 K in a CO--N$_2$ ice (Oberg et al.  2005), and 1180
K in a CO--H$_2$O ice (Collings et al. 2003; Fraser et al. 2004).  In
this study, we will adopt the value 885 K.

There is overwhelming observational evidence, mainly towards
pre-stellar cores, that N$_2$H$^+$ remains in the gas phase at larger
densities than CO (e.g. Caselli et al. 1999; Bergin et al.  2001;
Tafalla et al. 2004; Pagani et al. 2004; Crapsi et al. 2005).  Since
N$_2$H$^+$ is believed to be formed from N$_2$, these observations
suggest that N$_2$, which is the major nitrogen reservoir, freezes-out
onto grains at higher densities than CO.  The reason for that is not
fully understood, because the binding energy of N$_2$, measured in the
laboratories, is only slightly lower that the CO binding energy: 790 K
in a pure N$_2$ ice and 855 K in a N$_2$-CO ice (Oberg et al.  2005).
However, the effect of having N$_2$ in the gas phase where CO
molecules disappear greatly influences the abundances of H$_3^+$ and
its isotopomers, because it causes an additional term of H$_3^+$
(H$_2$D$^+$ and HD$_2^+$) destruction in Equation \ref{eq:ratio1} (and
Eqs. \ref{eq:ratio2} and \ref{eq:ratio3}).  Therefore, we decided to
adopt a semi-empirical approach, and to assume that the binding energy
of N$_2$ is 0.65 times that of CO, following the theoretical studies
of other authors (e.g. Bergin et al. 1995, 1997).  However, we will
discuss the effect of a larger N$_2$ binding energy, as measured in
laboratory, in \S \ref{hidden}.

We treat the freeze-out of CO and N$_2$ in a time-dependent way.  CO
molecules freeze out onto the grain mantles at a rate:
\begin{equation}
k_\mathrm{dep}=S \left<\pi a_{gr}^{2} n_\mathrm{g}\right> v_\mathrm{CO}
\end{equation}
where we adopted a sticking coefficient $S=0.3$ (Burke \& Hollenbach
1983), and a mean grain radius of $a_\mathrm{gr}$.  The grain number
density $n_{g}$ is given by the (mass) dust-to-gas ratio $\fduga$
multiplied by the gas density $n$, and divided by the grain mass (see
Eq.\eqref{eq:xgrain}).  A similar equation can be written for N$_2$.

Frozen CO and N$_2$ molecules can be released back into the gas phase
by thermal evaporation, and we followed the first order desorption
kinetics approach to describe it (Hasegawa \& Herbst 1993):
\begin{equation}\label{kev}
k_\mathrm{ev}=\nu_{0} \exp[-E_\mathrm{b}/kT]
\end{equation}
where $\nu_0=10^{12}$s$^{-1}$ is the lattice frequency, $E_{b}$ is the
binding energy per molecule of CO and N$_2$ ice respectively, and $k$
is the Boltzman constant.  

Cosmic rays also contribute to the release of CO and N$_2$ from the
ice, with a rate of (Hasegawa \& Herbst 1993):
\begin{equation}\label{kcr}
k_\mathrm{cr}=9.8\times10^{-15}\frac{\zeta}{3\times10^{17}s^{-1}} 
~\mathrm{s}^{-1} \quad.
\end{equation}

%% Cecilia: And lets also take this out here, OK?
%% The cosmic ray ionization rate $\zeta$ is of major importance for the
%% resulting deuteration of H$_3^+$: on the one hand, it regulates the
%% ultimate fraction of CO and N$_2$ in the gas phase, and, on the other
%% hand, it also influences the gas ionization degree, and, therefore the
%% overall abundance of H$_3$$^+$ and its isotopomers.  Thus, we will
%% throughly discuss the dependence of the results on the cosmic ray
%% ionization rate in \S \ref{major}.

The time dependent number density of gaseous CO, $n_\mathrm{CO}$, is
therefore given by the solution of the following equations:
\begin{eqnarray}
\frac{\mathrm{d}n_\mathrm{CO}}{\mathrm{d}t}
= -k_\mathrm{dep} n_\mathrm{CO} + (k_\mathrm{ev}+k_\mathrm{cr})
 \cdot n_\mathrm{CO}^\mathrm{ice}\\
  n_\mathrm{CO}^{\mathrm{ice}}+n_\mathrm{CO}= n_{\rm H_2}\cdot A_\mathrm{CO}
\end{eqnarray}
where we adopted the CO abundance $A_\mathrm{CO}$ observed in
molecular clouds, $9.5 \times 10^{-5}$ (Frerking et al. 1982).
$n_\mathrm{CO}^\mathrm{ice}$ is the number density of frozen CO
molecules.  Similar equations can be written for N$_2$, where the
N$_2$ abundance in molecular clouds is assumed to be $4\times10^{-5}$
(Bergin et al. 1995, 1997).  In our model, the CO and N$_2$ abundances
in the gas phase across the disk depend therefore on the time.  They
start at time=0 equal to their relative molecular cloud abundances.
At large times, they are given by the equilibrium between thermal
evaporation, cosmic ray desorption and freeze-out onto the mantles
(Leger et al. 1986).

Note that we do not treat in any way the CO and N$_2$
photo dissociation by the UV and/or X-rays photons emitted by the
central star.  Therefore, our model just describes the regions where
those photons are fully shielded, i.e.  the warm molecular layer and
the cold midplane of sufficiently massive disks.
As our model focuses
on the H$_3^+$ and its isotopomers only, it does not have any vocation
in treating the CO and/or N$_2$ chemistry, other than computing their
disappearance from the gas phase because of the freezing onto the dust
grains.  Besides, our model does not consider any possible grain
surface chemistry involving CO and N$_2$, which may transform (part
of) condensed CO (and N$_2$) into different, more complex molecules.

\subsection{Charge balance}
\label{sec:charge-balance}
Ionization in disks can be due to cosmic rays, X rays and UV rays.
For the current study we restrict ourselves to regions of the disk
which are shielded from UV radiation and X rays, i.e. to the disk
midplane.  In this case, cosmic rays are the dominating source of
ionization (Semenov, Wiebe \& Henning 2004).

In regions where the heavy-elements bearing molecules are depleted,
the positive charge is carried by the H-bearing species, namely
H$_3^+$, its isotopomers, and H$^+$.  The negative charge, on the
other hand, is mostly carried by free electrons.  The fraction of
electrons attached to grains remains small, in particular if the
smallest grains have been removed by coagulation (Walmsley et
al. 2004).  We will therefore ignore negatively charged grains when
computing the abundances of H$^+$, H$_3^+$ (with isotopomers), and
$e^-$.

Similar to H$_3^+$ (see \S \ref{sec:chem-deut-h_3+}) H$^+$ is formed
by the interaction of cosmic rays with H$_2$, with a rate $0.08\zeta$.
Recombination of H$^+$ occurs mainly on grains because the direct
recombination with electrons is too slow,
and the proton exchange with HD.
The abundance of H$^+$ is therefore independent of the electron
density.  Equating the formation and destruction rates, we have:
\begin{equation}\label{eq:H+}
x_\mathrm{H^+}=\frac{0.08\zeta}{
2k_\mathrm{HD}[{\rm D}]+
k_\mathrm{gr}x_\mathrm{gr}}
\end{equation}
where $\zeta$ is the cosmic ray ionization rate.
%% and $\beta$ is the
%% dissociative recombination rate, measured by McCallum et al. 2003 to
%% be $4\times10^{-7}$ at 10 K.  The above equation implicitly assumes
%% that the negative charge is carried by electrons, which very likely is
%% the case for the dense regions studied here.
%% and $k_{HD}=1.0\times10^{-9} e^{-464K/T}+e^{-634K/T}$ (Walmsley et al. 2004).
The recombination rate on grains k$_\mathrm{gr}$ is
discussed in \S~\ref{sec:recombination-grains}.

The electron abundance may be derived by equating the cosmic ray
ionization rates with the processes removing electrons from the gas.
%% The other important source of free electrons is the branch of H$_2$
%% ionization that leads to H$_3^+$, with a rate $\zeta$.  The negative
%% charge will be mostly carried by free electrons, while the total
%% number of electrons attached to grains remains small, in particular if
%% the smallest grains have been removed by coagulation (Walmsley et al. 2004).  For
%% simplicity, we therefore ignore the fraction of electrons on grains
%% and derive the electron abundance by considering the equilibrium of
%% cosmic ray ionization of H$_2$ leading to the formation of H$^+$ and
%% H$_3^+$, and recombination of H$^+$ on grains as well as of H$_3^+$ on
%% grains and with free electrons.
As source term of free electrons we consider the reactions
$\mathrm{H_2+cr\rightarrow H_3^++e^-}$ and $\mathrm{H_2+cr\rightarrow
H^++e^-}$. As electron sink term we use the recombination of
H$_3^+$ and its isotopes with electrons, with a dissociative
recombination rate $\beta$, measured by McCall et al. (2003) to be
$4\times10^{-7}$ at 10 K. In addition, we assume that every
recombination of H$_3^+$ (and its isotopomers) and H$^+$ on grains
remove an electron from the gas. This is a reasonable assumption because
positively charged grains quickly recombine with free electrons,
consistent with the very low abundance of positively charged grains
found by (Walmsley et al. 2004). 
This leads to the following balance equation for the electron density:
\begin{equation}
\label{eq:1}
\zeta (1.08) n_{\mathrm{H}_2} =
\left( \beta x_{\{\mathrm{H},\mathrm{D}\}_3^+} 
       + k_\mathrm{gr} x_{\mathrm{H}^+}
       + \frac{k_\mathrm{gr}}{\sqrt{m/m_\mathrm{H}}} 
       x_{\{\mathrm{H},\mathrm{D}\}_3^+}
\right) n_{H_2} n_{e^-}
\end{equation}
where $x_{\{\mathrm{H},\mathrm{D}\}_3^+}$ is the total abundance of
H$_3^+$ and its isotopomers, and $m$ is the average mass of these
molecules which we set equal to $4.5m_\mathrm{H}$. Together with
charge conservation $x_{\mathrm{e}^-} = x_{\mathrm{H}^+} + x_{H_3^+} +
x_{\rm H_2D^+} + x_{\rm D_2H^+} + x_{\rm D_3^+}$, a 2nd degree
equation for the electron density results which can be easily solved.

We can consider two limiting cases.  If the electron abundance is set
by the recombination of electrons with H$_3^+$ and its isotopes, we
can expect the canonical dependence
$x_{e^-}\propto\sqrt{\zeta/n_{H_2}}$.  If on the other hand the
electron abundance is set by recombination of H$^+$ on grains, the
electron abundance will be $x_{e^-}\propto \zeta/n_\mathrm{gr}$ where
$n_\mathrm{gr}$ is the number density of grains.  We will use this
difference below to identify the main recombination process in the
model calculations.

\subsection{Recombination on grains}
\label{sec:recombination-grains}

\begin{figure}[htbp]
%\centerline{\includegraphics[width=9.0cm]{../chemistry/output/jtildeerror.eps}}
\centerline{\includegraphics[width=9.0cm]{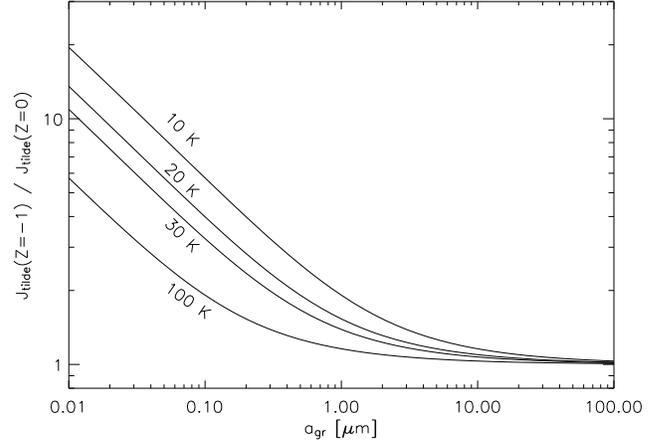}}
\caption{Ratio ${\tilde J}(Z_\mathrm{gr}=-1)/{\tilde
  J}(Z_\mathrm{gr}=0)$ as function of the grain
  radius, for different temperatures (see text).}
\label{fig:jtilde} 
\end{figure}
At low electron densities, collisions of ions with grains become an
important contributer to the recombination rates.  This is true in
particular for protons, and, depending on the rate coefficients for
dissociative recombination, also for molecular ions.  At high
densities, the charge distribution is dominated by neutral and singly
negatively charged grains.  Draine and Sutin (1987) argue that the
sticking coefficients for positive ions on both neutral and negatively
charged grains should be 1.  The different recombination rates then
only depend on the Coulomb focusing factor in the collision cross
section.  The Coulomb focusing factor ${\tilde J}$ is given by Draine
\& Sutin (1987) as a function of the grain charge, ion charge and the
gas temperature.  For a full treatment, the relative abundances of
neutral and negatively charged grains would have to be calculated, as
e.g. done by (Walmsley et al. 2004).  However, in the case of really dense
environments, in which the abundance of small grains is small due to
the effects of dust coagulation, a simplified treatment is possible.
Fig.\ref{fig:jtilde} shows the ratio ${\tilde
  J}(Z_\mathrm{gr}=-1)/{\tilde J}(Z_\mathrm{gr}=0)$.  While for very
small grain sizes and extremely low temperatures this ratio can be
very large, in the region of parameter space studied in this paper the
factor is reasonably small.  For 0.1\um, the ratio is already down to
5.7.  For larger grains and somewhat higher temperature, the ratio
gets closer to one.  For simplicity we therefore choose the following
approach.  We assume that a fraction $f_{-}$ of the grains is
negatively charged, and that $1-f_{-}$ of the grains are neutral.  The
recombination rate coefficient $k_{gr}$ is then given by (Draine \&
Sutin 1987):
\begin{equation}\label{eq:coulomb}
k_{gr}=\left( \frac{8 k T}{\pi m_H} \right)^{1/2} \pi a_{gr}^2
\left( f_{-} {\tilde J}(Z=-1)+(1-f_{-}){\tilde J}(Z=0) \right)
\end{equation}
Choosing $f_{-}=0.3$, we expect an error in the recombination rates
from this approach no larger than a factor of 3.

Finally, because of the uncertainties of the grain size distribution
in the midplane of a disk, we use only a single grain size
$a_\mathrm{gr}$ for each model.  Assuming a density of the grain
material of 2.5~gr~cm$^{-3}$, the dust-to-grain ratio in number density
is given by:
\begin{equation}\label{eq:xgrain}
x_\mathrm{gr}=3.2\times10^{-12}\left(\frac{\fduga}{0.01}\right)
\left(\frac{a_\mathrm{gr}}{0.1\mu{\rm m}}\right)^{-3}
\end{equation}

\subsection{H$_2$D$^+$ and D$_2$H$^+$ line intensities}\label{intensities}

H$_2$D$^+$ and D$_2$H$^+$ both appear in the para and ortho forms.
Gerlich, Herbst and Roueff (2002) have discussed in detail the issue,
and found that at low temperatures the ortho to para ratio of
H$_2$D$^+$ is close to unity. Subsequently, Walmsley et al. (2004)
have shown that the H$_2$D$^+$ ortho to para ratio indeed varies
weakly with the density in completely depleted regions, and it is
around 0.2 at densities $\geq 10^{6}$ cm$^{-3}$.  Since the present
study focuses on the outer disk midplane where the densities are
larger than $10^{6}$ cm$^{-3}$, we do not calculate the H$_2$D$^+$
ortho-to-para ratio from a chemical network, but simply assume it to
be 0.3.  Finally, Flower et al. (2004) find that the HD$_2^+$
ortho-to-para ratio is equal to 10 for large densities and low temperatures.

The ortho and para forms of both H$_2$D$^+$ and HD$_2^+$ have ground
transitions in the submillimeter to Far-IR (Tera-Hertz) range (Hirao
\& Amano 2003): 372.4 and 1370.1 GHz for the o-H$_2$D$^+$ and
p-H$_2$D$^+$ respectively, and 1476.6 and 691.7 GHz for the o-HD$_2^+$
and p-HD$_2^+$, respectively.  The o-H$_2$D$^+$ and p-HD$_2^+$
transitions are observable with ground based telescopes (CSO and JCMT
today, APEX and ALMA in the near future), whereas the other two
transitions can only be observed with airborne and satellite
telescopes, in particular SOFIA and the upcoming HERSCHEL mission.  In
this study, we report the line intensity of the four transitions,
computed with a non-LTE code which treats self-consistently the line
optical depth (Ceccarelli et al. 2003)\footnote{The code has been
  adapted to the disk geometry.}, assuming that the disk is seen
face-on.  Note that, for the disks studied in this article, the line
optical depth in the face-on configuration never exceeds unity though,
so that the lines are in practice always optically thin.  The
collisional coefficients are not well known, and we assumed that the
critical density of all the H$_2$D$^+$ and HD$_2^+$ ground transitions
are $10^{6}$ cm$^{-3}$ derived using the H$_2$D$^+$ scaled collisional
coefficients by Black et al. (1990).  Since the regions where the
lines originate have relatively large densities ($\geq 10^7$ cm$^-3$),
the levels are mostly LTE populated, and the uncertainty on the
collisional coefficients is, therefore, not of great importance here.
The line fluxes are given in erg s$^{-1}$cm$^{-2}$ assuming a source
distance of 140 pc.  For an easy comparison with the presently
available observations, we also give the o-H$_2$D$^+$ (at 372 GHz) and
p-HD$_2^+$ (at 691 GHz) line fluxes in main beam temperature, assuming
that the observations are carried out at the CSO and JCMT telescopes
respectively.

%%%%%%%%%%%%%%%%%%%%%%%%%%%%%%%%%%%%%%

\section{Results}\label{sec:results}

In this section we discuss the results of the model calculations,
namely the abundances of the H$_2$D$^+$, HD$_2^+$ and D$_3^+$ across
the disk.  One major goal of this article is to give predictions for
observations of the two deuterated forms of H$_3^+$ that have ground
rotational transitions in the submillimeter wavelength range:
H$_2$D$^+$ and D$_2$H$^+$. 

We first explore in detail our standard case, which is a disk with
canonical values for parameters like dust-to-gas ratio, cosmic ray
flux and grain sizes, properties not easily derived from observations.
In\S~\ref{major} we study the dependence on these parameters.  In
\S~\ref{minor} we address system properties that are generally known
from other observations: the total dust mass, the disk size, and the
stellar luminosity.  Finally, in \S~\ref{hidden} we discuss the
influence of one important model assumption, the binding energy of
N$_2$ and CO ice.

\subsection{The standard model}\label{sec:standard-model}

Our standard model is a disk with a dust disk mass equal to
$2\times10^{-4}$ M$_{\odot}$ surrounding a 0.5 L$_{\odot}$ star whose
T$_\mathrm{eff}$=3630 K (see \S 2.1).  The dust-to-gas ratio is 1/100
in mass, and we use the canonical cosmic ray ionization rate
$\zeta=3\times 10^{-17}$ s$^{-1}$.  We assume that the disk is
$10^{6}$ yr old, and extends up to 400 AU. Figure \ref{fig:struct}
shows the density and temperature profiles across the disk of our
standard case.

\begin{figure}[tb]
 \includegraphics[angle=0,width=1.\columnwidth]{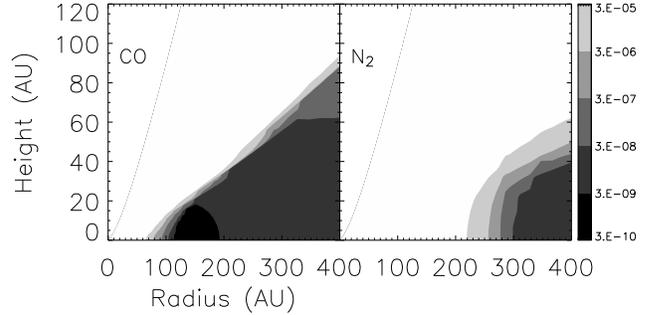}
 \caption{Abundance profiles of CO (right panel) and N$_2$ (left
 panel) across the disk, for our standard case. The different colors show
 the abundance as given by the scale bar.}
\label{fig:n2co}
\end{figure}

Figure \ref{fig:n2co} shows the abundances of CO and N$_2$
respectively, across the disk. CO molecules disappear from the gas
phase where the dust temperature is around 25 K, at a radius larger
than about 100 AU and at heights lower than about 1/4 of the
radius. Since the density is large in those cold regions, the drop in
the abundance is very sharp, more than a factor 100.  The largest
CO depletion occurs between about 100 and 200 AU, where it is larger
than $10^4$ times. At larger radii it is a factor 10 lower. N$_2$,
however, remains in the gas phase in a much larger region, and,
therefore, dominates the destruction rates of H$_3^+$ (and its
isotopomers). It is only in the very outer regions, at a radius larger
than about 300 AU that N$_2$ freezes-out onto the grains and almost
totally disappears from the gas phase.  The intermediate region, where
CO is depleted but N$_2$ is not, is where the deuteration of H$_3^+$
takes place in a ``moderate'' form, namely where the
H$_2$D$^+$/H$_3^+$ ratio is enhanced with respect to the elemental D/H
abundance, but D$_3^+$ is not the dominant positive charge carrier.

\begin{figure}[tb]
\includegraphics[angle=0,width=1.\columnwidth]{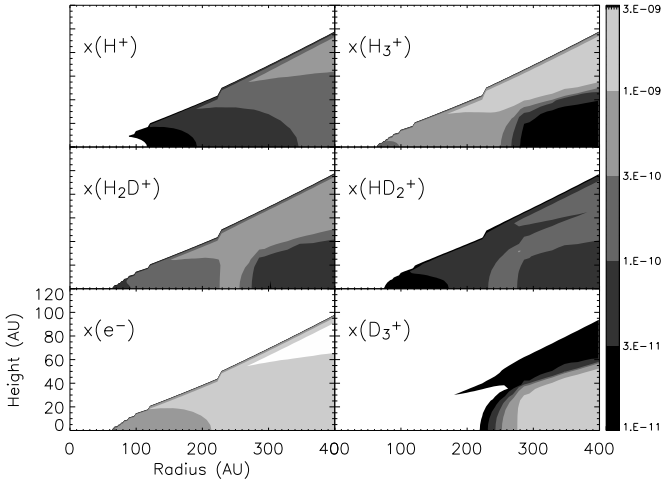}
 \caption{ Chemical structure of the standard disk: fractional
   abundances with respect to H$_2$ as function of the disk radius and
   height, in AU.  Top left panel: H$_3^+$.  Top right panel: H$^+$.
   Center left panel: H$_2$D$^+$. Center right panel: HD$_2^+$.
   Bottom left panel: electrons.  Bottom right panel: D$_3^+$.  The
   color levels show the abundance as given by the scale bar.}
\label{chemi}
\end{figure}

Figure \ref{chemi} shows the H$^+$, H$_3^+$ isotopomers, and electron
abundance profiles of the standard disk respectively.  We limit the
plot of the chemical composition to regions where the CO depletion is
larger than a factor of 3. The choice of 3 is a somewhat
arbitrary.  It represents when the CO abundance is similar to that of
N$_2$, or, in other words, when the role of these two molecules
-believed to be the most abundant gaseous molecules at low
temperatures- becomes of similar importance in the H$_3^+$ reactions,
the H$_3^+$ deuteration becomes to be significant, and our simplified
model correct.
The electron abundance, is between 3 and $30\times10^{-10}$ in the
outermost region, at radius $\geq 50$ AU, increasing going towards
the upper layers of the disk. The results are in reasonable agreement
with the ionization structure computed by Semenov et al. (2004). In
most of the disk, the positive charge is carried by H$_3^+$, which
becomes less important where also the N$_2$ disappears from the gas
phase, at radius $\geq$300 AU.  H$_2$D$^+$ is the most abundant
H$_3^+$ isotope, except in the radius $\geq$300 AU region, where
D$_3^+$ takes over, and become the most abundant positive ion. H$^+$
never plays a major role as charge carrier across the entire disk,
even though its abundance is comparable to that of H$_2$D$^+$ in outer
upper levels of the disk.

The average column densities of a face-on disk of the H$_3^+$ isotopes
and the electrons in the region where the CO depletion is larger than
3 are reported in Table~\ref{coldens}.  The D$_3^+$ column density is
the largest, and it is about $30\%$ larger that of H$_2$D$^+$. The two
isotopes together account for almost half of the overall positive
charge in the disk midplane. The HD$_2^+$ column density is twice
smaller than the H$_2$D$^+$ column density. These values are dominated
by the outer region, where both CO and N$_2$ are frozen-out onto the
grains, and hence, D$_3^+$ becomes the dominant positive charge
carrier.
\begin{table}[tb]
\caption{Column densities averages across the face-on disk of the
  H$_3^+$ isotopes and electrons in the standard disk model.}
\label{coldens}
\begin{center}
\begin{tabular}{lc}
%\hline
 Species    & Column  Density (cm$^{-2}$)\\\hline
 H$_2$D$^+$ & $1.6\times10^{13}$ \\
 HD$_2^+$   & $5.1\times10^{12}$ \\
 D$_3^+$    & $3.3\times10^{13}$ \\
 e$^{-}$    & $1.0\times10^{14}$ \\
%\hline
\end{tabular}
\end{center}
\end{table}

The line intensities of the four ortho and para H$_2$D$^+$ and the
HD$_2^+$ ground transitions are reported in Table~\ref{fluxStandard}.
\begin{table}[hb]
\label{fluxStandard}
\begin{tabular}{l|cccc}
%\hline
                      & o-H$_2$D$^+$      & p-H$_2$D$^+$ & o-HD$_2^+$ & p-HD$_2^+$ \\
\hline
Transition            & $1_{1,0}-1_{1,1}$ & $1_{0,1}-0_{0,0}$ 
                      & $1_{1,1}-0_{0,0}$ & $1_{1,0}-1_{0,1}$                      \\
$\nu$ (GHz)           & 372.4             & 1370.1       & 1476.6     & 691.7      \\
Flux erg/s/cm$^{2}$   & 8.2e-18           & 1.0e-16      & 5.7e-17    & 2.6e-18    \\
T$_{mb}\Delta$v (mK km/s)& 18.8$^a$           & -            & -          & 4.8$^a$      \\
%T$_{mb}$ (mK)         & 25.8                & -            & -          & 12    \\ %N2=2e-5 
%T$_{mb}\Delta$v (mK)         & 5.6                & -            & -          & 6.8    \\ %N2=0
%\hline
\end{tabular}\newline
$^a$ Note: the main beam efficiency is assumed to be 0.6 at CSO and 0.3 at JCMT.
\caption{Line fluxes of the ground transitions of the ortho and para
form of H$_2$D$^+$ and the HD$_2^+$ respectively, for the standard
case.  The velocity-integrated line intensities, expressed in main
beam temperatures, T$_{mb}\Delta$v are computed assuming observations at CSO
and JCMT of the o-H$_2$D$^+$ and p-HD$_2^+$ transitions respectively.}
\label{fluxStandard}
\end{table}

\subsection{Ionization rate, dust-to-gas ratio and grain sizes as
  important factors for the abundances} \label{major} 

Important parameters of the model are the cosmic ray
ionization rate, the dust-to-gas ratio, and the grain sizes, all of
which are relatively ill known.  One goal of this theoretical work is
to provide predictions of observable quantities which can help to
constrain these parameters, which all play a key role in the evolution
of the proto-planetary disks and the eventual planet formation (see
Introduction).

\begin{figure}[tb]
\includegraphics[angle=0,width=0.8\columnwidth]{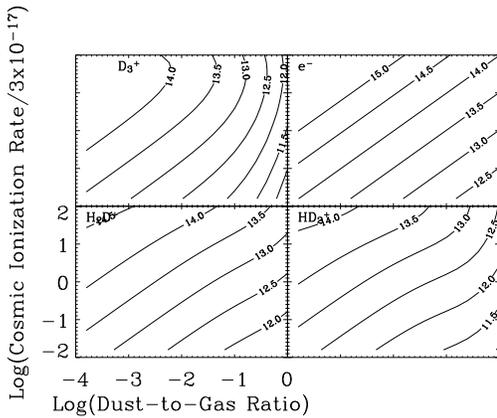}
 \caption{Logarithm of the column density averaged over the disk
surface of D$_3^+$ (upper left panel), H$_2$D$^+$ (lower left panel),
HD$_2^+$ (lower right panel) and electrons (upper right panel). The
plots have been obtained for our standard case, namely with a grain
radius of 0.1 $\mu$m.}
\label{fig:coldens1}
\end{figure}

In order to investigate the influence of the cosmic ray flux and of
the dust-to-gas ratio, we have computed a grid of models where we
vary these parameters within a factor 100 of the standard value.
Since the structure of the disk is fully determined by the dust
opacity, changing the dust-to-gas ratio simply implies scaling the gas
density while all other factors like gas temperature and grain number
density remain constant.  Figure~\ref{fig:coldens1} shows the column
densities as a function of the cosmic ray ionization rate normalized to
$3\times10^{-17}$, and of the dust-to-gas ratio.  It is immediately
obvious that the electron column density varies as a simple powerlaw
$\propto \sqrt{\zeta/\fduga} \propto \sqrt{\zeta n_{H_2}}$.  This
corresponds to an electron abundance dependence
$x_{e^-}\propto\sqrt{\zeta/n_{H_2}}$, indicating that cosmic ray
ionization and dissociative recombination of electrons with the
charge carrying molecules (H$_3^+$ and its isotopes in this case) are
determining the electron abundance.  This is consistent with the
isotopes of H$_3^+$ being the main charge carriers.  Because
throughout the plot, the dust grain density $n_\mathrm{gr}$ is
constant (changing \fduga{} only changes the gas density), recombination
on grains would lead to a dependence $x_{e^-}\propto\zeta/n_{H_2}$.

The shape of the contours for the column densities of H$_2$D$^+$ and
D$_2$H$^+$ are similar to those of e$^-$, again indicating that these
molecules are the dominant charge carriers.  D$_3^+$ shows a similar
behavior for very low dust-to-gas ratios (i.e. high gas densities)
and normal or low cosmic ray ionization rates.  However, for very low
gas densities (high dust-to-gas ratios) and for very high ionization
rates, the contours become vertical, indicating that the column
density becomes independent of the ionization rate. A similar but
smaller effect can also be observed for the column densities of
H$_2$D$^+$ and HD$_2^+$ at high dust-to-gas ratios. This is due to
the fact that at low gas densities (corresponding also to low
electron densities), the recombination of these molecules on grains
starts to become important and leads to a faster decrease in the
abundances.

\begin{figure}[tb]
\includegraphics[angle=0,width=0.8\columnwidth]{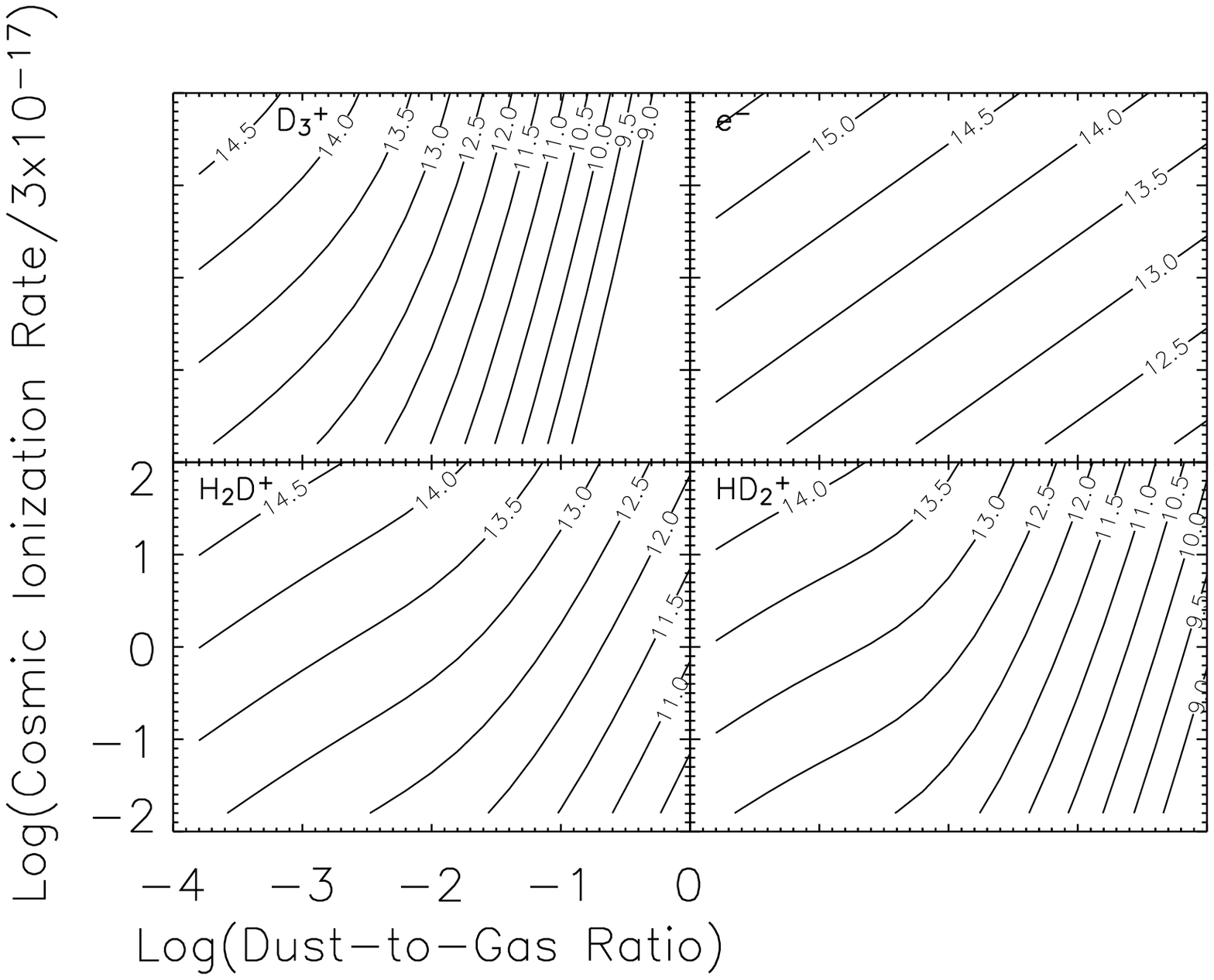}
 \caption{Same as Figure \ref{fig:coldens1} but for a grain radius of
 0.01 $\mu$m.}
\label{fig:coldens2}
\end{figure}
\begin{figure}[tb]
\includegraphics[angle=0,width=0.8\columnwidth]{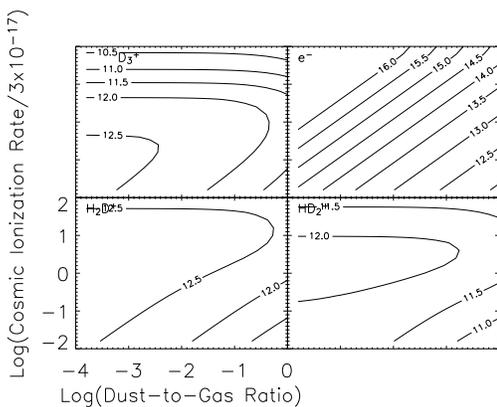}
 \caption{Same as Figure \ref{fig:coldens1} but for a grain radius of
 1.0 $\mu$m.}
\label{fig:coldens3}
\end{figure}

Figures \ref{fig:coldens2} to \ref{fig:coldens3} are similar to
Fig.~\ref{fig:coldens1}, but using a disk model with different grains
sizes.  We have used grain sizes of 0.01\um, and 1.0\um, to study the
cases of very little dust growth compared to the interstellar medium,
and significant dust growth.  Since the total dust mass in the model
is constant, reducing the grain size corresponds to an increase in the
number density of grains, and also in the total grain surface per
cm$^3$.  This does favor recombination on grains as a destruction
route of H$_3^+$ and its isotopes.  This is indeed visible, in
particular at high dust-to-gas ratios, the column densities of all
three molecules are strongly decreased, and almost independent of the
ionization rate.

A different effect can be observed in the calculation for large
grains.  In this case, the column density of electrons greatly
exceeds that of the H$_3^+$ isotopes, in particular in the upper left
part of the diagram, at high ionization rates and high gas densities.
The electron column density contours are closer together, indicating a
linear dependence on ionization rate and gas density.  As discussed in
\S~\ref{sec:charge-balance}, this is an indication that H$^+$ is now
the dominating positive charge carrier and that recombination of
H$^+$ on grains has taken over as main electron destruction
mechanism.  The reason for the significant decrease in deuterated
H$_3^+$ abundance compared to cases with smaller grains is caused by
smaller CO/N$_2$ depletions.  The larger grains provide less grain
surface for these molecules to freeze out and shift the abundance ratios
away from the deuterated species towards H$_3^+$.  The column
densities of H$_2$D$^+$, HD$_2^+$ and D$_3^+$ all have a maximum at
normal cosmic ray ionization rates.  If the cosmic ray flux is further
increased, the column densities degrease again, because the CO/N$_2$
depletion becomes less efficient.  Also for smaller grains, the same
effect is expected, but at much higher fluxes.

\begin{figure}[tb]
\includegraphics[angle=0,width=0.8\columnwidth]{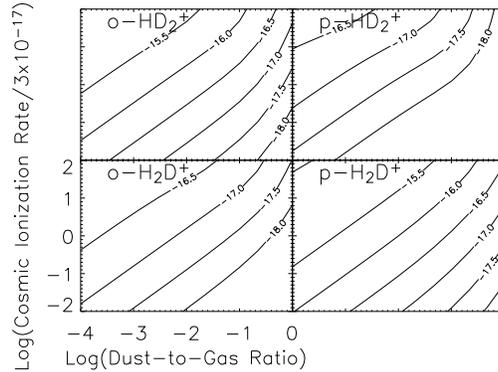}
 \caption{Logarithm of the line fluxes as function of the dust-to-gas
ratio and the cosmic ray ionization rate. The fluxes are in
erg/s/cm$^2$ and computed for a source at 140 pc of distance. The four
panels show the four observable transitions: o-H$_2$D$^+$ (372 GHz),
p-H$_2$D$^+$ (1370 GHz), o-HD$_2^+$ (1476 GHz) and p-HD$_2^+$ (691
GHz).}
\label{fig:fluxes}
\end{figure}
\begin{figure}[tb]
\includegraphics[angle=0,width=1.\columnwidth]
%         {../chemistry/output/Tmb_paper.ps}
         {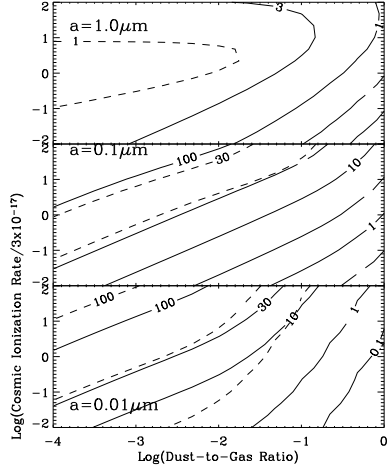}
\caption{Velocity-integrated line intensities, expressed as mean beam
temperature, of the o-HD$_2^+$ $1_{1,0}-1_{1,1}$ (solid lines) and
p-HD$_2^+$ $1_{1,0}-1_{0,1}$ (dashed lines) transitions for CSO and
JCMT observations respectively.  The contour plots are in mK km/s. Upper,
center and lower panels refer to different grain radii: 0.01, 0.1 and
1.0 $\mu$m respectively.}
\label{tmb-major}
\end{figure}

Figure~\ref{fig:fluxes} shows the dependence of the H$_2$D$^+$ and
HD$_2^+$ line fluxes on the cosmic ray ionization rate and the
dust-to-gas ratio.  For an easy comparison with observations
obtainable at CSO and JCMT we also show plots with the
velocity-integrated mean beam temperatures of the o-HD$_2^+$ at 372 GHz and
p-HD$_2^+$ at 691 GHz lines respectively (Figure~\ref{tmb-major}).
The curves of the o-H$_2$D$^+$ and p-HD$_2^+$ line intensities run
practically parallel, so that the simultaneous observation of the two
lines would not depend on the parameters of the model, but the
ortho-to-para ratio of each species. They can therefore be ``safely''
used to derive the relative ortho-to-para ratio of H$_2$D$^+$ with respect
to HD$_2^+$.

\subsection{The basic stellar and disk properties}\label{minor}

The basic stellar and disk parameters such as stellar luminosity, dust
mass in the disk and disk size can normally derived independently.  We
therefore do not enter into a full parameter study of these
quantities.  Instead, we limit ourselves to describe the main trends
by looking at single-parameter changes from the standard model.
Table~\ref{tab:stellar-params} summarizes the results of these
calculations.  The stellar and disk parameters mainly change the local
densities in the disk, or the temperature.  Temperature increase
influences the excitation of the lines, and may reduce depletion.
Density changes modify the degree of ionization and again the
depletion efficiency.  The results, summarized in
Table~\ref{tab:stellar-params}, show that changes of a factor of two
in any of the parameters have only limited effects on the line fluxes,
in all cases less than a factor of two.
The largest changes occur when the disk radius is changed, mostly because
of the larger/smaller integration emitting region.
\begin{table}[htb]
\begin{center}
\begin{tabular}[tb]{l|cccc}
%\hline
Parameter                 & o-H$_2$D$^+$  & p-H$_2$D$^+$ & o-HD$_2^+$ & p-HD$_2^+$ \\
\hline
$L_{\star}\times2$        & 1.21 & 1.46 & 1.23 & 0.86 \\ 
$L_{\star}/2$             & 0.73 & 0.73 & 0.73 & 0.77 \\
$M_\mathrm{dust}\times2$  & 1.44 & 1.45 & 1.46 & 1.46 \\
$M_\mathrm{dust}/2$       & 0.75 & 0.82 & 0.73 & 0.60 \\
$R_\mathrm{disk}\times2$  & 1.64 & 1.43 & 1.38 & 1.65 \\
$R_\mathrm{disk}/2$       & 0.51 & 0.73 & 0.65 & 0.40 \\
%\hline
\end{tabular}
\end{center}
\caption{\label{tab:stellar-params} Changes of the line fluxes emitted
  by the disk for different values of the stellar and disk parameters.
  The numbers are the fluxes of the four ground ortho and para
  H$_2$D$^+$ and HD$_2^+$ lines relative to the standard case.}
\end{table}

%%%%%%%%%%%%%%%%%%%%%%%%%%%%%%%%%%%%%%%%%%%%%%%%%
\subsection{High N$_2$ binding energy}\label{hidden}

As mentioned in \S \ref{sec:co-n_2-freezing}, there is a discrepancy
between the measured binding energy of N$_2$ onto the grains -which is
similar to the CO binding energy- and the observations in pre-stellar
cores -which show that N$_2$ molecules condense onto the grains at a
larger density, and therefore, suggest a N$_2$ binding energy lower
than the CO one.  In our standard model, discussed in the previous
paragraphs, we adopted that the N$_2$ binding energy is 0.65 times
that of CO, in agreement with the modeling of the pre-stellar cores
observations (e.g. Bergin et al. 1995, 1997).  It is also worth
noticing that observations of N$_2$H$^+$ in a proto-planetary disk
support indeed the lower binding energy for N$_2$ (Qi et al. 2003).
However, for the sake of completeness, here we explore the influence
of this assumption on the results.

As shown in Figure \ref{fig:n2co} and previously discussed, the low
N$_2$ binding energy implies that N$_2$ molecules remain in the gas
phase in a large region where CO is frozen-out onto the grains.  It is
therefore clear that the presence or not of N$_2$ in the gas phase
will have a strong impact on the modeling results.

\begin{figure}[tb]
\includegraphics[angle=0,width=0.8\columnwidth]{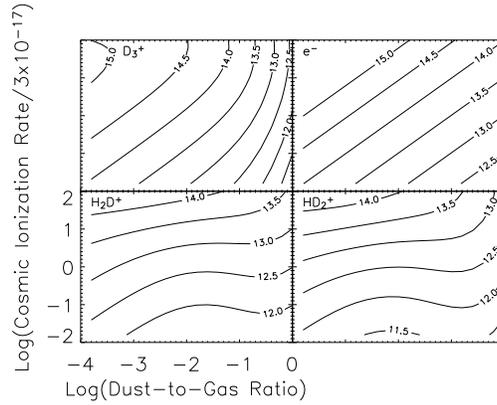}
 \caption{Logarithm of the column density averaged over the disk
surface of D$_3^+$ (upper left panel), H$_2$D$^+$ (lower left panel),
HD$_2^+$ (lower right panel) and electrons (upper right panel).These
plots have been obtained assuming that the N$_2$ binding energy is
equal to that of CO.}
\label{fig:coldens_noN2}
\end{figure}

Figure \ref{fig:coldens_noN2} shows the column densities of H$_3^+$,
H$_2$D$^+$, HD$_2^+$ and e$^-$ as function of the dust-to-gas ratio
and the cosmic ionization ratio, similarly to Figure
\ref{fig:coldens1}.  The comparison of the two figures speaks for
itself.  If N$_2$ has a binding energy comparable to that of CO
(Figure \ref{fig:coldens_noN2}), D$_3^+$ is the positive charge
carrier across most of the outer disk, in the standard case
(dust-to-dust ratio equal to 0.01, and cosmic ray ionization rate
equal to $3\times 10^{-17}$ s$^{-1}$), and H$_2$D$^+$ and HD$_2^+$ are
30 times less abundant.  On the contrary, if N$_2$ has a reduced
binding energy (Figure \ref{fig:coldens1}) the positive charge is
almost equally shared between D$_3^+$, H$_2$D$^+$ and H$_3^+$.  This is
because the N$_2$ collisions are an important destroyer of the
H$_3^+$, H$_2$D$^+$ and HD$_2^+$, preventing the ``total conversion''
of H$_3^+$ into D$_3^+$.

If the CO and N$_2$ binding energies are similar, increasing and/or
decreasing the dust-to-gas ratio does not change much the H$_2$D$^+$
and/or HD$_2^+$ column densities. In practice, increasing/decreasing
the gas density with respect to the dust density does not change the
H$_2$D$^+$ and HD$_2^+$ column densities, but just increases/decreases
the D$_3^+$ column density -the positive charge carrier. This has an
important observational consequence, for the H$_2$D$^+$ and/or
HD$_2^+$ observed column densities cannot constrain the dust-to-gas
ratio in a large interval.
However, it is worth noticing that, in principle, the ratio of the
derived average column densities can help to discriminate whether
N$_2$ is present in the gas phase or not.  Finally, observations of
the N$_2$H$^+$ in proto-planetary disks will also help to clarify
whether N$_2$ remains in gas phase after CO condenses (Qi et al. 2003).

%%%%%%%%%%%%%%%%%%%%%%%%%%%%%%%%%%%%%%
\section{Discussion}\label{sec:discussion}

\subsection{Simplified treatment of the chemical network}

In this paper we have introduced a simplified chemical network to
compute the abundances of deuterated H$_3^+$.  The network treats
only the four isotopes of H$_3^+$, electrons, H$^+$, CO, and N$_2$.
Comparison with more extensive networks shows that the abundances of
the \{H,D\}$_3^+$ ions can be calculated accurately under the conditions of
significant CO and N$_2$ depletion.  The reason for this success is that
the chemistry indeed becomes very simple under these conditions.  CO
and N$_2$ are the last heavy-elements bearing molecules with significant
abundances to leave the gas phase and to freeze out on dust grains.
Together with dust grains, they act as the main destroyers of H$_3^+$
and its isotopes.  The simplified chemical network is, under these
conditions, in fact complete in that it treats all important species
and reactions.  Under the assumption that the freeze-out of CO and
N$_2$ has reached its steady state, the steady state solution for this
network can be derived directly, no time integration of the rate
equations is necessary.  The chemical network, therefore, lends itself
for large parameter studies of cold gas.

\subsection{The main charge carrier in the disk midplane}

In the discovery paper of the H$_2$D$^+$ line in DM Tau and TW Hya,
Ceccarelli et al (2004) had suggested that H$_2$D$^+$ may be the main
charge carrier under the conditions found in a disk, which would allow
for a direct derivation of the degree of the column density of
positive ions by observing H$_2$D$^+$.  With an H$_2$ column density
derived from dust continuum observations, this number directly leads
to the degree of ionization, a much sought-after property of disk gas
because of its importance for the magneto-rotational instability, the
likely driver of disk turbulence.  This conclusion has to be softened
in the light of the present results.  Our calculations show that
H$_2$D$^+$, HD$_2^+$, D$_3^+$ and H$_3^+$ all carry similar amounts of
charge, indicating that the observed column densities of H$_2$D$^+$
and HD$_2^+$ provide both an order-of-magnitude estimate of the degree
of ionization, and a lower limit.  It was shown earlier (Walmsley et
al. 2004) that complete depletion in protostellar cores generally
leaves H$^+$ as the main charge carrier.  This trend, however, clearly
ends at the densities found in disks: H$_3^+$ and its isotopes
contribute equally or dominate the charge balance.

In this context, the binding energy of N$_2$ is of critical
importance.  If N$_2$ freezes out at the same moment as CO, then the
disk midplane are indeed fully depleted of heavy element bearing
molecules.  Then, there is no important destruction route for D$_3^+$
anymore, and the abundance of D$_3^+$ sores to dominate the charge
balance. If N$_2$ indeed stays in the gas longer than CO, as suggested
by observations in a proto-planetary disk (Qi et al. 2003),
H$_2$D$^+$ and HD$_2^+$ are produced in significant abundances and do
provide a measure of the degree of ionization.

\subsection{H$_2$D$^+$ and HD$_2^+$ as tracers of cold gas}

One major question we aim to answer with the current study is: under
what conditions are the observable lines of H$_2$D$^+$ and HD$_2^+$
tracers of the gas mass in the disk midplane?  For these molecules to
be a tracer, the observed line intensities must be a strong function
of the dust-to-gas ratio.  As we have seen in
Figures~\ref{fig:coldens1} and \ref{fig:coldens_noN2}, the answer to
this question depends on the behavior of N$_2$.  If N$_2$ freezes-out
at lower temperatures than CO -as supported by the observations in
a proto-planetary disk (Qi et al. 2003)-, then the observed line
intensities do depend strongly on the dust-to-gas ratio and can be
used to measure it.  However, under the conditions of complete
depletion, i.e. if also N$_2$ is completely removed from the midplane
gas, the line intensities become insensitive to the dust-to-gas ratio.

Fortunately, observations of N$_2$H$^+$ in disks can
discriminate what is the reality, namely whether in the disk midplane
CO and N$_2$ disappear simultaneously from the gas phase or not, and,
therefore, whether H$_2$D$^+$ and HD$_2^+$ are the main positive
charge carriers or D$_3^+$ is.  In the first case, observations of
H$_2$D$^+$ and HD$_2^+$ will give a measure of the dust-to-gas ratio,
once the cosmic ionization rate is known.  This also implies that
H$_2$D$^+$ and HD$_2^+$ measure the ionization degree in the disk
midplane too, as assumed by Ceccarelli et al (2004).  Once again,
we emphasize that the available observations (Qi et al. 2003) support
this case.

\subsection{Dust coagulation}
An important result of this study is the dependence of the chemical
structure of the disk midplane on the average dust grain size.  Figure
\ref{tmb-major} shows that the line intensities of the o-H$_2$D$^+$
and p-HD$_2^+$ are strongly affected by the average grain sizes, if
they are larger than 0.1 $\mu$m radius.  In practice, if small dust
grains are significantly depleted in the midplane, the line
intensities decrease dramatically, because CO and N$_2$ do not
freeze-out efficiently on the reduced grain surfaces (see the
discussion in \S \ref{major}), and no substantial molecular
deuteration takes place.  The detection of the o-H$_2$D$^+$ and
p-HD$_2^+$ at the level of the current instrumental sensitivity is,
therefore an indication that enough grain surface is still available
to account for the molecular depletion.  Two scenarios are possible
for this.  First, coagulation might not be efficient or aggregate
destruction is efficient in keeping the number density of small grains
at reasonable values.  This is in fact in agreement with recent
results of grain coagulation which suggest that coagulation at maximum
efficiency is inconsistent with the observed opacities and shapes of
flaring circumstellar disks (Dullemond \& Dominik 2005).  Another
possibility is that the depletion of CO and N$_2$ is very efficient in
the beginning, before coagulation happens.  Efficient coagulation
could then form large bodies that bury the ice inside and protect it
from cosmic ray desorption.  These considerations show that detailed
models of coagulation, vertical mixing and freeze-out in combination
with observations of H$_3^+$ isotopes can be useful diagnostics of
grain coagulation.

%%%%%%%%%%%%%%%%%%%%%%%%%%%%%%%%%%%%%
\section{Conclusions}\label{sec:conclusions}

We have computed the chemical structure of the midplane of disks
surrounding solar type protostars.  We have described the derived
abundances and the resulting line intensities of the deuterated forms
of H$_3^+$ in a standard case, and for a wide range of values of the
dust-to-gas ratio and cosmic ray ionization rate. We have also discussed
in detail the influence of the average grain sizes on the
results. Finally, we have reported values also for different stellar
luminosities, disk masses and radii, and discussed in detail the case
of the N$_2$ freezing onto the grain mantles simultaneously with
CO.  Our main conclusions are:

\begin{itemize}
\item H$_3^+$ deuteration is significant in the midplane of
  proto-planetary disks around solar type protostars.  In our standard
  case, the positive charge is carried by H$_3^+$ in a large zone of
  the disk midplane, except at the very outer radii, larger than about
  300 AU, where D$_3^+$ takes over. H$_2$D$^+$ is the most abundant
  H$_3^+$ isotope across most of the disk, at radii less than about
  300 AU.
\item Contrary to what was earlier assumed, H$_2$D$^+$ is {\it not}
  the dominating positive ion in the standard case.  H$_3^+$
  and its isotopes are equally important.  With increasing depletion
  of CO and N$_2$, D$_3^+$ becomes more abundant, eventually
  dominating as the positive charge carrier.  Also, H$^+$ is only
  of minor importance.  While at the pre-stellar-core densities, lower
  than 10$^7$cm$^{-3}$, H$^+$ dominates (Roberts et al. 2003, 2004;
  Walmsely et al. 2004), in the disk midplane this is no longer the
  case. 
\item The midplane chemical structure and the predicted line
   intensities of the H$_2$D$^+$ and HD$_2^+$ are a strong function of
   the local cosmic ray ionization rate, which regulates the overall
   ionization degree, and the CO and N$_2$ depletion across the disk.
 \item The chemical structure and line intensities are also sensitive
   to the dust-to-gas ratio as long as depletion of heavy-element
   bearing molecules is not complete. This happens if the N$_2$
   binding energy is lower than the CO binding energy, as observed in
   pre-stellar-cores and possibly in proto-stellar disks.  If,
   on the contrary, N$_2$ freezes-out simultaneously with CO, as
   laboratory experiments would rather suggest, since the major charge
   carrier is D$_3^+$, increasing/decreasing the dust-to-gas ratio
   does not change appreciably the H$_2$D$^+$ and HD$_2$ column
   densities and/or line intensities.
\item The grain size has strong influence on the line strength
   expected for H$_2$D$^+$ and D$_2$H$^+$.  Small grains accelerate
   recombination and reduce the abundances of all positive ions, and
   therefore decrease the degree of ionization.
\end{itemize}

This article focuses entirely on theoretical predictions.  A
forthcoming paper (Dominik et al., in preparation) will analyze in
detail the case of DM Tau, where the o-H$_2$D$^+$ line at 372 GHz has
been detected, applying the model here developed to a practical case.
Likely, with the advent of ALMA, with its great sensitivity and
spatial resolution, observations of both the o-H$_2$D$^+$ at 372 GHz
and p-HD$_2^+$ at 691 GHz will be possible on a routine/systematic
base and with spatial resolution.  Those observations promise to be
very fruitful, and to bring unique information on the physical status
of the midplane of the disks surrounding solar type protostars, likely
similar to the progenitor of our own Solar System.  Besides,
observations of p-H$_2$D$^+$ and o-HD$_2^+$ by out-of-the-atmosphere
instruments will hopefully measure the actual ortho-to-para ratio of
these species.  Finally, D$_3^+$ would be likely the best diagnostics
for the gas mass and ionization in low mass disks, but the only
possibility to observe it is by absorption at $\sim$5
$\mu$m (Ramanlal \& Tennyson 2004; Flower et al. 2004). This requires
observations towards disks almost edge-on, but not completely, for the
absorption is against the inner dust continuum, which must therefore
be detectable. Such disks are hard to find, but new observations by,
for example, SPITZER may well discover such sources, and these
absorption observations may be possible in the future.

\acknowledgements{It is a pleasure to thank Paola Caselli, Charlotte
  Vastel and Malcolm Walmsley for comments on the manuscript, and Kees
  Dullemond for his collaboration on constructing the disk model.  We
  also wish to thank an anonymous referee for careful reading the
  manuscript. We acknowledge Travel support through the Dutch/French
  van Gogh program, project VGP 78-387.}

%%%%%%%%%%%%%%%%%%%%%%%%%%%%%%%%%%%%%%%%%%%%%%%%%%%%%%%%%%%%%%%%%%%%%%%%%%%% 
{}

\end{document}